\documentclass{INTERSPEECH2023}
\usepackage{subcaption}

\interspeechcameraready

\title{Language-universal phonetic encoder for low-resource speech recognition}
\name{Siyuan Feng, Ming Tu, Rui Xia, Chuanzeng Huang, Yuxuan Wang}
\address{
  Speech and Music Intelligence (SAMI), ByteDance}
\email{\{fengsiyuan.ee,mingtu,rui.xia,huangchuanzeng,wangyuxuan.11\}@bytedance.com}

\begin{document}

\maketitle
 
\begin{abstract}
Multilingual training is effective in improving low-resource ASR, which may partially be explained by phonetic representation sharing  between languages. In end-to-end (E2E) ASR systems, graphemes are often used as basic modeling units, however graphemes may not be  ideal for multilingual phonetic sharing. In this paper, we leverage International Phonetic Alphabet (IPA)  based language-universal phonetic  model to improve low-resource ASR performances, for  the first time within the attention encoder-decoder architecture.  We propose an adaptation method on the phonetic IPA model to further improve the proposed approach on extreme low-resource languages. Experiments carried out on the open-source MLS corpus and our internal databases show   our approach outperforms baseline monolingual models and most   state-of-the-art works. Our main approach and adaptation are effective on extremely low-resource languages,  even within domain- and language-mismatched scenarios.





\end{abstract}



\section{Introduction}
\label{sec:intro}
Recently, end-to-end automatic speech recognition (ASR) systems have achieved remarkable performances in high-resource languages e.g. English \cite{baevski2020wav2vec,hsu2021hubert}. The availability of huge amounts of training data and the increase  of computational capabilities are two essential driving forces.
There are around 7000 languages in the world \cite{speech2020scharenborg}. Most of the languages are considered under-resourced, i.e., having limited speech and/or text resources. This poses  a major challenge in developing high-performance
ASR systems for low-resource languages.

Low-resource ASR has been an active research topic recently \cite{Huang2013cross,toshniwal2018multilingual,li2021scaling,swietojanski2012unsupervised,hou2021exploiting,conneau2020unsupervised}. One mainstream research line is multilingual training  \cite{Huang2013cross,toshniwal2018multilingual,li2021scaling}, i.e. pooling together training speech and transcripts of multiple languages to train a single ASR model. 
A second research line is crosslingual transfer learning \cite{swietojanski2012unsupervised,hou2021exploiting}, i.e., leveraging an ASR model trained by   non-target, usually resource-rich  language(s) as the seed model, retraining it with a target low-resource language's data.
A third research line is self-supervised learning (SSL)   \cite{oord2018cpc,baevski2020wav2vec,hsu2021hubert,chung2021w2v,baevski2022data2vec}, typically starting with unsupervised pretraining followed by supervised finetuning. SSL  methods are effective  especially when  a large amount of untranscribed speech is available.
In addition, data augmentation and self-training,  which share the 
idea of increasing the supervised data amount, were studied for low-resource ASR \cite{meng2021mixspeech,zhang2021xlst}.

This paper follows  the multilingual research line. The success of multilingual training for low-resource ASR suggests that
the superiority of multilingual models over monolingual ones is presumably explained by (1) more training data; and (2)  the ability of sharing linguistic (or more specifically, phonetic) knowledge  across languages \cite{Zelasko2020That_interspeech}.
In this work, we are more interested in 
the second perspective.  
As different languages vary greatly in terms of fundamental units (phonemes), orthography, phonotactics etc, we argue that how to enable and facilitate phonetic representation sharing  is not a trivial problem. 
In the era of hybrid deep neural network hidden Markov model (DNN-HMM)    architecture,   a predominant multilingual approach is  sharing  hidden layers between languages meanwhile keeping output layers language specific \cite{Huang2013cross}. With this idea,  phonetic information sharing is realized within  hidden layers but not in output layers of the DNN model. 
In the  E2E ASR paradigm, multilingual training is usually realized by 
directly pooling  data of all the languages as the training data, and merging together graphemes of these languages as the model's vocabulary
\cite{li2021scaling,toshniwal2018multilingual,kim2018towards}. 
While the use of graphemes as basic modeling units brings simplicity and has been proven effective, it could suffer from the data sparsity issue   especially in the multilingual setting \cite{li2019bytes}. Moreover, graphemes are not accurate descriptors of speech pronunciations in some languages (e.g. English), which might adversely affect multilingual phonetic sharing.

Past works explored   basic modeling units other than graphemes for effective multilingual training in E2E ASR. 
In \cite{li2019bytes}, the authors proposed to use bytes  \cite{gillick2015multilingual}. In \cite{liu2022hierarchical_hoffmann}, Huffman code was adopted. In \cite{pratap2020massively,zhou2018multilingual}, subword units were adopted.
The use of language-universal International Phonetic Alphabet (IPA) \cite{international1999handbook_ipa} symbols as basic modeling units  was   investigated  in the E2E architectures including 
connectionist-temporal-classification (CTC) \cite{tong2017multilingual},  E2E lattice-free maximum mutual information (LF-MMI) \cite{tong2019investigation} and self-supervised pretraining \cite{Feng2023ssl_ipa}.
On a different but relevant task of phone recognition,  \cite{Zelasko2020That_interspeech} analyzed the efficacy of IPA based basic units for 
multilingual phonetic sharing.
They  observed huge improvements
of phone recognition
from monolingual IPA models to the multilingual IPA model, especially on low-resource languages.
Follow-up analyses   \cite{feng2021how_phonotactics,zelasko2022discovering} compared model architectures between the listen-attend-spell (LAS) \cite{chan2016listen} and the DNN-HMM  for IPA based phonetic sharing,
and found that the LAS architecture is more effective.



From \cite{Zelasko2020That_interspeech,feng2021how_phonotactics,zelasko2022discovering}, it is clear that  IPA is an effective means to facilitate multilingual phonetic representation sharing, particularly in the LAS model architecture. 
On the other hand, while there were studies  on  IPA based multilingual training in CTC and LF-MMI architectures \cite{tong2017multilingual,tong2019investigation}, the use of IPA in the LAS-based E2E models for ASR tasks has not been studied.
Motivated by this, we attempt to leverage multilingual, hopefully language-universal phonetic representations learned by the LAS-based  IPA model \cite{Zelasko2020That_interspeech}, to improve low-resource ASR.
Concisely, we propose to firstly train a multilingual IPA model with IPA transcribed speech, then finetune the IPA model with a target language's orthographically transcribed speech. An adaptation operation is optionally applied to the trained IPA model, in order to strengthen its
phonetic representation learning of extremely low-resource languages, hence  improve the ASR performance of these languages.



\section{Proposed approach}
\label{sec:approach}

The proposed approach consists of two main stages, which will be described in Sections \ref{subsec:approach_ipa_model} and \ref{subsec:approach_asr_finetune} respectively.  The optional adaptation method will be discussed in \ref{subsubsec:approach_ipa_model_adaptation}.


\subsection{Language-universal IPA model}
\label{subsec:approach_ipa_model}

IPA is a standardized representation of speech sounds in written form  \cite{international1999handbook_ipa}. It is independent of  languages. This makes the IPA system intrinsically suitable for multilingual phonetic sharing.
An IPA model   can recognize a speech utterance into a sequence of speech  sounds  symbolized by the IPA system.
To train a multilingual IPA model,    $N$ distinct languages' supervised training data (denoted as $\mathcal{F}$) is required.
Grapheme-to-IPA conversion is  applied  to convert orthographic transcriptions of every language into phonetic IPA transcriptions.
The vocabulary of the multilingual IPA model, i.e. the  output layer of the model, consists of the union of IPA symbols present in the $N$ languages. 
We treat modifier symbols, such as   long vowels [\textlengthmark] as separate basic units following \cite{Zelasko2020That_interspeech}.
For instance, the sound [\textipa{a}\textlengthmark] is recognized as two consecutive tokens [\textipa{a}], [\textlengthmark]. This enables sharing between [\textipa{a}] and [\textipa{a}\textlengthmark].
Such sharing would not 
happen if [\textipa{a}\textlengthmark] were treated as a whole unit.

The multilingual IPA model learns a phonetic representation that is (quasi) language universal: fundamental speech sounds from  different languages that share the same IPA symbol could be pronounced the same, or slightly differently, and they are mapped to the same output of the model. 
The fact that the language-universal IPA model captures a broad range of languages' phonetic information is preferred in  subsequent ASR finetuning, especially for languages with limited data.
Intuitively, increasing $N$ and making languages in $\mathcal{F}$ more diverse help the IPA model   approach language universality.

The LAS based E2E ASR architecture is adopted for 
developing the IPA model. The model consists of an encoder and a decoder. The language-universal phonetic representations are expected to be  mainly modeled by the encoder of the IPA model. 



\subsubsection{Adaptation to an extremely low-resource language}
\label{subsubsec:approach_ipa_model_adaptation}
In practice, it may well be that the language-universal IPA model's training data amounts of different languages are (highly) imbalanced. Phonetic representations of languages with relatively less data are at risk of being underfit by the IPA model, compared to phonetics of languages with more data. We hypothesize that this could result in sub-optimal ASR performance when finetuning the IPA model to an extremely low-resource language.
To address this issue, we propose an adaptation operation on the IPA model. 
Given the unadapted, well-trained language-universal IPA model, targeting at a particular language,  we utilize this language's IPA transcribed training data to retrain the model for a few epochs with a small learning rate. With such an operation, compared to the unadapted IPA model,  the adapted IPA model is  strengthened in capturing the phonetic representation of the target language, meanwhile  without losing language universality. 


\subsection{Target-language ASR finetuning}
\label{subsec:approach_asr_finetune}
A language-universal IPA model  (either unadapted or adapted) is taken as the seed model. The IPA model's encoder is kept,  while its  decoder is replaced by a randomly initialized one of the same layer size and number. 
The seed model is finetuned with a target language's speech and orthographic transcripts. The vocabulary of the  finetuned ASR model consists of the byte-pair encoding (BPE) tokens present in  BPE-tokenized transcripts of the target language.

In principle,   ASR finetuning could also be done on  the whole language-universal IPA model, instead of only on the encoder part. However,  the decoder  of the LAS-based IPA model heavily captures  phonotactics information of the training languages, and according to \cite{feng2021how_phonotactics},  phonotactics of non-target languages would hurt the recognition performance  on a target language. Therefore, in this work we focus on finetuning only   the encoder of the language-universal IPA model.

\section{Experimental setup}

\label{sec:setup}
\subsection{Databases and evaluation metric}
\label{subsec:exp_databases}
Speech corpora used for training the language-universal IPA models include the open-source  Multilingual Speech (MLS) corpus \cite{pratap2020mls} and our internal data. 
The MLS corpus covers Polish (PL), Portuguese (PT), Italian (IT), Spanish (SP), French (FR), Dutch (DU), German (GE) and English (EN), all derived from read audiobooks. 
The amounts of training, development and test data per language
are listed in Table \ref{tab:database}. Our internal data consists of EN and Japanese (JP), both derived from the video domain. The total  hours of  EN and JP internal data are 12.3k and 8.8k respectively. 
We consider the following multilingual training sets for training  language-universal IPA models:
\textbf{MLS-7}: we merge  training sets of PL, PT, IT, SP, FR, DU and GE, summing up to \textbf{6.0k} hours; 
\textbf{MLS-8}: 
    we merge  \textbf{MLS-7} with the EN training set, summing up to \textbf{50k} hours. 
\textbf{NT-2}: we merge  our  EN and JP internal datasets, summing up to \textbf{21.1k} hours.
\textbf{}


Data used for  target-language ASR finetuning is taken from MLS only.
As this paper focuses primarily on  low-resource ASR, and 
the EN training set size in the MLS corpus is an order of magnitude larger than the other languages' training sets, we do not carry out  English ASR finetuning.
For each of the 7 MLS languages, the full training set and a  training subset of 100 hours  are  prepared for finetuning respectively. The 100-hour training subset is randomly chosen from the full training set of a language.


The finetuned ASR model for a target language is evaluated on the language's test set in the MLS corpus, using word error rate (WER) as the metric.
Throughout our experiments, the development data partition is used exclusively for  monitoring the  training process.

The relations of the multilingual training sets and MLS test sets are summarized in Table \ref{tab:domain_languages}. The column `Domain matched' denotes if the training set and the test sets are from the same domain, and the column `Languages' denotes 
the existence of
 training languages covered (target, T) and not covered (non-target, N) by the test sets.

\begin{table}[!t]
\renewcommand\arraystretch{1}
\centering
\caption{The sizes of the MLS training, development and test sets in hours for every language, and the number of IPA symbols covered by every language.}
\resizebox{  0.9 \linewidth}{!}{%
\begin{tabular}{c|cccccccc}      
\toprule
Language & PL & PT & IT & SP & FR & DU & GE & EN \\
\midrule
Train & 104 & 161  & 247  & 918  & 1.1k & 1.6k & 2.0k & 44.7k\\
Dev & 2.1 & 3.6 &5.2 & 10.0 & 10.1 &12.8 &14.3 & 15.8\\
Test & 2.1 &3.7 &5.3 &10.0 &10.1 &12.8 &14.3 &15.6 \\
\midrule
\# IPA symbols & 35& 37&  29&  31&  45&  40&  47&  48\\
\bottomrule
\end{tabular}%

}
\label{tab:database}
\end{table}

\begin{table}[!t]
\renewcommand\arraystretch{1}
\centering
\caption{Summary of   domain match/mismatch between training and test data, and existence of target/non-target languages in the training data.}
\resizebox{  0.6 \linewidth}{!}{%
\begin{tabular}{c|cc}      
\toprule
Training set  & Domain matched & Languages \\
\midrule
MLS-7 & Y & T  \\
MLS-8 & Y & T\&N \\
NT-2 & N & N \\
MLS-7\&NT-2 & Y\&N & T\&N\\
\bottomrule
\end{tabular}%

}
\label{tab:domain_languages}
\end{table}

\subsection{Language-universal IPA model}
\label{subsec:setup_ipa}
An open-source LanguageNet software \cite{hasegawa2020grapheme} is utilized to convert orthographic transcripts of all the languages used in  the experiments into   phonetic IPA transcripts. The number of IPA symbols covered by every language in the MLS corpus is listed in Table \ref{tab:database}.  The numbers of IPA symbols covered by the \textit{MLS-7} and \textit{MLS-8} sets, i.e. union sizes of IPA symbols covered by the multiple languages,  are 87 and 95 respectively. For the EN and JP internal data, their numbers of IPA symbols are 48 and 22. The numbers of IPA symbols covered by  \textit{NT-2} and \textit{MLS-7\&NT-2}  are 55 and 96.


The IPA model consists of a
Conformer \cite{guo2021recent_conformer} encoder and a Transformer \cite{karita2019comparative} decoder.
It is implemented using ESPnet \cite{watanabe2018espnet}.
The encoder and decoder have 18  and 2 layers respectively, with 4 attention heads,   768 attention dimensions and 2048 position-wise feed forward (PFF) dimensions. 
The  encoder contains a 2-layer CNN with a kernel size of 31. 
Multiple  IPA models are trained, each  using  one of the multilingual training sets in \{MLS-7, MLS-8, NT-2, MLS-7\&NT-2\}.  
In every training experiment, the model is trained for 60 epochs  with joint CTC and attention objectives and the CTC weight is 0.1, using the  Adam optimizer \cite{kingma2014adam}, a peak learning rate of 0.001 and a warm-up step size of 2000. The final model is obtained by averaging   over models of epochs 51 to 60.

\subsection{IPA model adaptation}
Adaptation is performed on the language-universal IPA model trained using the MLS-8 training set (denoted as the \textit{unadapted IPA model}). Adaptation is carried out once for a target language  in the  MLS corpus (excluding EN). The unadapted IPA model is retrained  with the  learning rate of 
$5\times 10^{-5}$ for a fixed 2 epochs.
 The  the learning rate and  epochs are determined based on our pilot experiments.


\subsection{Target-language ASR finetuning}
\label{subsec:setup_finetune}
ASR finetuning is performed using one language's  training set in MLS (excluding EN). 
ASR finetuning used BPE-tokenized orthographic transcripts. 
The BPE tokenization was implemented by    SentencePiece\footnote{https://github.com/google/sentencepiece}. The number of BPE tokens for each language is 5000. 
Prior to finetuning, the IPA model's decoder is randomly initialized, and its output layer is replaced with a new one, which corresponds to the BPE tokens present in the target language. The
number of training epochs, learning rate, warm-up size,    objective function and model averaging   follow  the settings in IPA model training (see Section \ref{subsec:setup_ipa}).
Throughout this work, a language model is not used.


\subsection{Baseline monolingual ASR}
Monolingual ASR models are trained and serve as the baseline. All the monolingual models take the same architecture as the ASR model described in Section \ref{subsec:setup_finetune}.
For every target language in the MLS corpus (excluding EN), speech and BPE-tokenized transcripts of the training set are used to train a monolingual model from scratch. The implementation of BPE tokenization keeps the same as that in Section \ref{subsec:setup_finetune}.
The number of training epochs, learning rate, warm-up size,  objective function and model averaging   follow  the settings in Sections \ref{subsec:setup_ipa} and \ref{subsec:setup_finetune}. In other words, a baseline monolingual ASR model for a target
language and the proposed model finetuned from the language-universal IPA model to the target language 
differ only in the initialization of the model's encoder. 

\section{Results and discussion}
\label{sec:results}
\subsection{Main results}
Word error rate (WER) results of the proposed approach, the  baseline monolingual models and state-of-the-art works on the MLS test sets are listed in Table \ref{tab:results_main}. 
\begin{table}[!t]
\renewcommand\arraystretch{1}
\centering
\caption{WER$\%$ of the proposed approach, baseline monolingual models and state of the arts on MLS test sets. The number enclosed in brackets denotes the number of parameters in a model. }
\resizebox{  \linewidth}{!}{%
\begin{tabular}{l|lllllll|c}      
\toprule
  & PL & PT & IT & SP & FR & DU & GE & Avg.  \\
\midrule
MLS monolingual \cite{pratap2020mls} & 21.66 & 20.52 &  11.78 & 6.68 &   6.58 &  13.09 &  7.10 & 12.49  \\
\quad + 5-gram LM \cite{pratap2020mls} & 20.39& 19.49& 10.54& 6.07& 5.58& 12.02& 6.49 & 11.51  \\
XLSR-53 (300M) + 4-gram LM \cite{conneau2020unsupervised} & 17.2& 14.7& 10.4&  6.3& 7.6& 10.8& 7.0& 10.6  \\
B0 (15 lang. init.; 370M) \cite{li2021scaling} & 10.9&  15.5& 10.1& 4.7& 6.1& 11.1& 5.0& 9.1 \\
E3 (15 lang. init.; 1B) \cite{li2021scaling} & 10.4 & 15.2 & 8.8 & 4.2 & 4.9& 9.9& 4.3& 8.2\\
JUST (600M) \cite{bai2022joint} &6.6&8.0&8.2&3.7&5.2&9.5&4.1&6.5 \\
\midrule
\multicolumn{9}{l}{Proposed approach and monolingual baseline finetuned with full training data}  \\
\midrule
Baseline monolingual (216M) & 15.93& 24.85& 14.00& 6.25& 6.02& 12.86& 7.25& 12.45 \\
 MLS-8 (216M) &  6.98& 12.71& 10.43& 4.84& 4.51& 10.80& 6.34& 8.09 \\
 \quad + Adaptation & 6.84& 12.48& 10.39& 5.04& 4.69& 10.82& 6.34& 8.09 \\
 MLS-7 (216M) & 14.14& 14.84& 10.89& 5.36& 4.97& 11.37& 6.90& 9.78 \\
 NT-2 (216M) & 9.56& 16.80& 11.77& 6.74& 5.12& 12.49& 7.00& 9.93 \\
 MLS-7\&NT-2  (216M) & 7.38& 13.64& 10.53& 5.00& 4.91& 11.44& 6.60& 8.50 \\
\midrule
\multicolumn{9}{l}{Proposed approach and XLSR-53 finetuned with 100-hour training data}\\
\midrule 
XLSR-53 (300M) + 4-gram LM \cite{conneau2020unsupervised} & 18.9& 15.7& 12.0& 7.9& 9.8& 10.9& 7.4& 11.8  \\
MLS-8 (216M) & 7.00 & 13.17 & 11.93 & 7.96 & 7.58 & 14.38 & 9.49 & 10.22   \\
 MLS-7 (216M) & 16.41 & 15.54 & 12.00 & 7.97 & 8.62 & 15.62 & 10.45 & 12.37\\
\bottomrule
\end{tabular}%
}
\label{tab:results_main}
\end{table}
The proposed approach, irrespective of using any of the  \textit{MLS-7}, \textit{MLS-8} and \textit{NT-2} multilingual training sets for training the IPA model, performs significantly better than the baseline monolingual models and the MLS official monolingual models \cite{pratap2020mls} in terms of the averaged WER. Notably, while the  monolingual models by  \cite{pratap2020mls} and by ours are implemented differently (wav2letter++ \cite{pratap2019wav2letter++} v.s. ESPnet),   with different training objectives (CTC v.s. hybrid CTC/attention), the averaged WER numbers  are close (12.49\% v.s. 12.45\%).
Our best system, i.e. finetuning from the IPA model trained by the MLS-8 set, achieves 4.4\% absolute WER reduction compared to the baseline monolingual system.
Looking at WER break-down to each target language, a positive correlation is found between the absolute WER reduction (from  monolingual to multilingual) of a language and the amount of training data of that language: Polish and Portuguese benefit  the most and German has  the smallest WER improvement.
The results demonstrate the effectiveness of language-universal phonetic representations learned by the IPA model encoder for improving ASR performances, particularly for languages with very limited training data.

Table \ref{tab:results_main} shows that, the best result achieved by the proposed approach performs better than XLSR-53 \cite{conneau2020unsupervised}, B0 and E3  \cite{li2021scaling}. 
Our approach does  not perform better than the JUST \cite{bai2022joint} except on the French set.
The XLSR-53 utilizes 56k hours of unsupervised data  (including but not limted to MLS) for pretraining and a 5-gram LM during decoding. The B0 and E3 models use 359k hours of training data covering 15 languages to train a seed model, followed by using the MLS  training data to finetune the ASR model. E3 has a larger model size than all the other  models in Table \ref{tab:results_main}. 
The JUST model uses the MLS full training sets for joint unsupervised and supervised training. The superior results of JUST seem to suggest that the integration of self-supervised  and supervised training strategies is a promising research direction, which we leave for future study.

Table \ref{tab:results_main} lists  WER results of the proposed approach and XLSR-53 by using only 100 hours of data per language in ASR finetuning. 
It should be noted that  the 100-hour finetuning sets used for  XLSR-53    are not identical to the sets used in our experiments, since the 100-hour subset information could  not be found in  \cite{conneau2020unsupervised}.
In the 100-hour finetuning  scenario, our approach taking the  IPA model trained with the MLS-8 set   performs better  than XLSR-53. 





\subsection{Impact of adaptation}
WER results of the proposed approach by applying adaptation are listed in Table \ref{tab:results_main}, in the row ``+ Adaptation''. 
The adaptation operation contributes to WER improvements  on Polish,   Portuguese and Italian, the three languages containing the smallest amounts of training data (100 $\sim$ 250 hours). 
For German and Dutch, the two languages having the largest amounts of training data, adaptation results in little to no difference.  For Spanish and French, adaptation leads to  WER degradation. 
Based on the WER results, 
we conclude that the proposed adaptation operation has a positive impact on target languages with extremely low-resource languages e.g. less than 250 hours. 
Suggested by the results, for languages with more training data, the proposed adaptation is expected to be  less or not necessary.





\subsection{Discussion about non-target languages and domain mismatch}
In the proposed approach, the comparison  between adopting  MLS-8 and MLS-7 as the IPA model training data  in Table \ref{tab:results_main} shows that including the non-target language  (English) in training a language-universal IPA model  is helpful to  target languages' ASR tasks. Moreover, this benefit  is seen on all  the seven target languages. Thus  a non-target language's  phonetic knowledge information is helpful in the proposed approach.

The system by adopting the NT-2 set for IPA model training achieves an averaged WER better than the baseline monolingual models (see Table \ref{tab:results_main}). As the NT-2 training set  is from a different domain than that of the MLS test sets, and is composed of non-target languages only (English and Japanese), the results indicate the proposed approach is effective even on the domain- and language-mismatched scenario.
On the other hand, the system with the NT-2 training set performs worse than   the    systems with MLS-7, MLS-8 and MLS-7\&NT-2. This is in line with expectation: MLS-7, MLS-8 and MLS-7\&NT-2 all contain domain-matched data of target languages. 
We observe an improvement by using the MLS-7\&NT-2 training set in IPA model training, comparing to the  systems with MLS-7 and with NT-2. This can be explained by a more language-universal phonetic representation learned in the system MLS-7\&NT-2. 
Lastly, the system MLS-7\&NT-2 underperforms the system MLS-8, which is probably due to the domain mismatch issue.



\subsection{Visualization of IPA model encoder representations}
\label{subsec:results_visualization}
\begin{figure}
    \centering
     \begin{subfigure}[b]{0.49\linewidth}
         \centering
         \includegraphics[width=\linewidth]{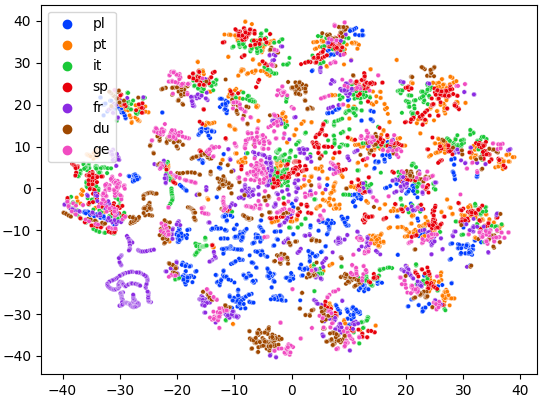}
         \caption{MLS-7 IPA model}
         \label{subfig:tsne_encoder_representations_mls7}
     \end{subfigure}
     \hfill
     \begin{subfigure}[b]{0.49\linewidth}
         \centering
         \includegraphics[width=\linewidth]{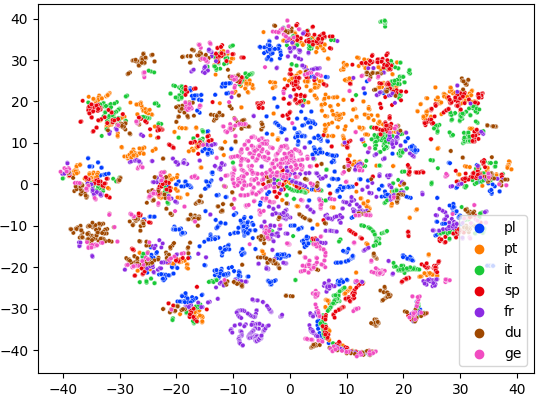}
         \caption{MLS-8 IPA model}
         \label{subfig:tsne_encoder_representations_mls8}
     \end{subfigure}
     \hfill
     \begin{subfigure}[b]{0.49\linewidth}
         \centering
         \includegraphics[width=\linewidth]{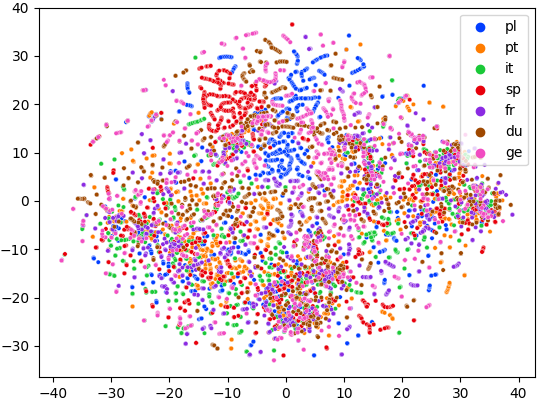}
         \caption{NT-2 IPA model}
         \label{subfig:tsne_encoder_representations_tn2}
     \end{subfigure}
     \begin{subfigure}[b]{0.49\linewidth}
         \centering
         \includegraphics[width=\linewidth]{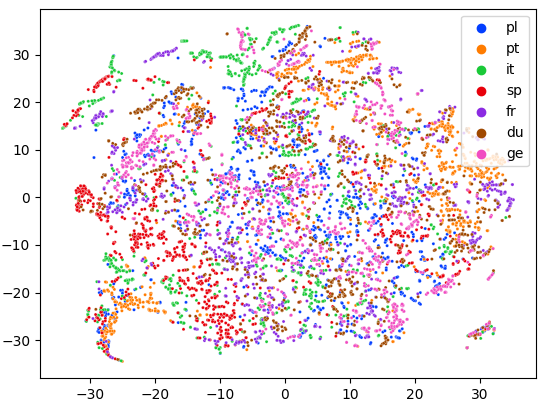}
         \caption{XLSR-53}
         \label{subfig:tsne_encoder_representations_xlsr-53}
     \end{subfigure}
        \caption{T-SNE visualizations on encoder representations of IPA models and the XLSR-53  \cite{conneau2020unsupervised} Transformer representation. }
        \label{fig:tsne_encoder_representations}
\end{figure}

To gain a deeper understanding on the phonetic representations learned by the IPA model, we apply t-SNE \cite{maaten2008visualizing} on the encoder output of several IPA models, and illustrate  visualization results  in 
Figures \ref{subfig:tsne_encoder_representations_mls7}, \ref{subfig:tsne_encoder_representations_mls8} and \ref{subfig:tsne_encoder_representations_tn2}.
An illustration result on the open-source XLSR-53's Transformer output representation is shown in Figure \ref{subfig:tsne_encoder_representations_xlsr-53}.
In one figure, every sample point stands for a speech frame, and the color denotes the language. These speech frames are randomly chosen from the MLS test sets,  1000 frames per language.  The selected frames are fixed for all the IPA models and the XLSR-53.
From Figures \ref{subfig:tsne_encoder_representations_mls7} and \ref{subfig:tsne_encoder_representations_mls8} we observe distinct phone-like patterns consisting of samples in  different colors. This indicates the two  language-universal IPA models, MLS-7 and MLS-8, are able to group together speech sounds of different languages that share the same or similar pronunciations. In comparison, from Figure \ref{subfig:tsne_encoder_representations_xlsr-53}, the representation learned by XLSR-53 is much less evident on phone-like patterns.  From Figure \ref{subfig:tsne_encoder_representations_tn2}, we observe that for the IPA model trained with the domain- and language-mismatched NT-2 set, 
clusters of samples in different colors exist to a mild extent.



\section{Conclusions}
This paper presented an approach to improving low-resource ASR, which leverages language-universal phonetic representations learned by an IPA model. An optional adaptation operation was proposed on the IPA model, to strengthen the IPA model's ability of capturing extremely low-resource languages' phonetic representations.
Experiments carried out on the   MLS corpus and our internal databases showed   the proposed approach outperforms baseline monolingual models and most of state-of-the-art works. Our main approach and adaptation are effective particularly on extremely low-resource languages. The approach brings improvements to the monolingual baseline even with  domain- and language-mismatched training data. Visualizations of IPA models' learned representations further confirmed the IPA model is capable of capturing language-universal phonetic representations.





\bibliographystyle{IEEEtran}
\bibliography{mybib}

\end{document}